\documentclass[aip,reprint,jcp]{revtex4-1}
\usepackage{graphicx}
\usepackage{subfigure}
\usepackage{adjustbox}
\usepackage{tikz}
\usetikzlibrary{calc,fadings,decorations.pathreplacing}

\usepackage[utf8]{inputenc}
\usepackage[T1]{fontenc}

\begin{document}
\title{Inelastic Charge Transfer Dynamics in Donor-Bridge-Acceptor Systems Using Optimal \
Modes}
\author{Xunmo Yang}
%\email[]{Your e-mail address}
%\homepage[]{Your web page}
%\thanks{}
%\altaffiliation{}
\affiliation{Department of Chemistry, University of Houston, Houston TX 77204}
\author{Eric R. Bittner}
\email[]{bittner@uh.edu}
\homepage[]{k2.chem.uh.edu}
%\thanks{}
%\altaffiliation{}
\affiliation{Department of Chemistry, University of Houston, Houston TX 77204}

\date{\today}

\begin{abstract}
We present a novel {\em ab initio} approach for computing intramolecular charge and
energy transfer rates based upon a projection operator scheme
that parses out specific internal nuclear motions that accompany
the electronic transition.
Our approach concentrates the coupling between the electronic and nuclear
degrees of freedom into a small number of reduced harmonic modes that can be
written as linear combinations of the vibrational normal modes of the molecular system
about a given electronic minima.
Using a time-convolutionless master-equation approach, parameterized
by accurate quantum-chemical methods, we  benchmark the approach
against  experimental results and predictions from Marcus theory for triplet energy transfer for a
series of donor-bridge-acceptor systems.
We find that  using only a single reduced mode--termed the ``primary''  mode,
 one obtains an accurate evaluation of the golden-rule rate constant and
insight into the nuclear motions responsible for coupling the initial and final electronic states.
We demonstrate the utility of the approach by computing the
inelastic electronic transition rates in a model donor-bridge-acceptor complex
that has been experimentally shown
that its exciton transfer pathway can be radically modified by mode-specific
infrared excitation of its vibrational mode.
\end{abstract}

%\pacs{}% insert suggested PACS numbers in braces on next line

\maketitle %\maketitle must follow title, authors, abstract and \pacs

%\chapterauthor{Xunmo Yang and Eric R Bittner\\
%University of Houston, Department of Chemistry}
%\chapter{Inelastic Charge Transfer Dynamics in Donor-Bridge-Acceptor Systems Using Optimal Modes}%

\section{Introduction}

\begin{figure}[h]
\includegraphics[width=0.5\columnwidth]{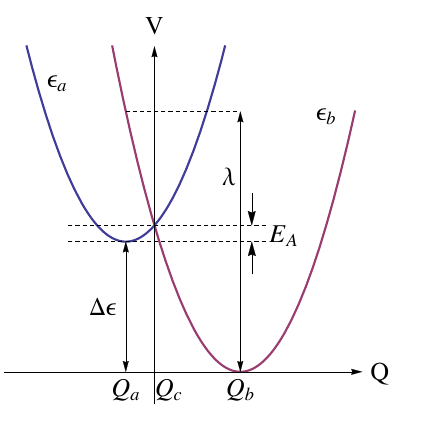}
\caption{Sketch of Marcus parabolas for a model energy or charge transfer system.
Labeled are the key parameters used to compute the Marcus rate constant (Eq. \ref{eq:marcus}).
Energies are given in eV and the collective nuclear displacement  is dimensionless.
}
\label{marcus}
\end{figure}

Energy and electronic transport plays a central role in a
wide range of  chemical  and biological systems.
It is the fundamental mechanism for transporting the  energy  of an
absorbed photon to a reaction center in light harvesting systems
and for initiating a wide range of photo-induced  chemical processes,
including vision, DNA mutation, and pigmentation.
The seminal model for calculating electron transfer rates was developed by
Marcus in the 1950's\cite{marcus1956theory,marcus1965theory,marcus1993electron}.
\begin{equation}
k_{Marcus}=\frac{2\pi}{\hbar}|V_{ab}|^{2}\frac{1}{\sqrt{4\text{\ensuremath{\pi}}k_{B}T\lambda}}e^{-(\lambda+ \Delta\epsilon)^{2}/4\text{\ensuremath{\lambda}}k_{B}T}.\label{eq:marcus}
\end{equation}
where $\lambda$ is  energy required to reorganize the environment
following the transfer of an electron from donor to acceptor.
and $\Delta \epsilon$ is the driving force for the reaction, as illustrated in Fig.~\ref{marcus}. If we assume that the nuclear motions about the equilibrium configurations of the
donor and acceptor species is harmonic,  the chemical reactions resulting from
energy or charge transfer events can be understood in terms of intersecting
diabatic potentials as sketched.    The upper and lower
curves are the adiabatic potential energy surfaces describing the nuclear dynamics
resulting from an energy or charge transfer event, taking the geometry of the donor
state as the origin.

%-------FIGURE 1

As the transfer occurs by crossing an energy barrier,
the transfer rate can be expected to be in the Arrhenius form
\begin{eqnarray}
k\propto e^{-E_{A}/k_{B}T},
\end{eqnarray}
with $E_{A}$ as the activation energy.
 Using $E_{A}={(\lambda+\Delta \epsilon)^{2}}/{4\lambda}$
 we can relate the activation energy to both the reorganization energy and driving force, $-\Delta \epsilon$.
One of the most profound predictions of the theory is that  as the driving force increases,
the transfer rate reaches a maximum and further increases in the driving force
lead to lower reaction rates, termed the inverted regime. The existence of the
inverted region was demonstrated unequivocally by Miller {\em et al.} \cite{miller1984intramolecular}
in an elegant series of experiments that systematically tuned the driving force, reorganization energy,
and diabatic coupling  by careful chemical modification of the donor and acceptor.

A number of years ago,  our group developed
a time-convolutionless  master equation approach for computing
state-to-state rates in which the coupling between  states depends upon the
nuclear coordinates\cite{pereverzev2006time}. This approach incorporates a
 fully quantum-mechanical treatment of both the nuclear and electronic degrees of freedom and recovers the
well-known Marcus expression in the semiclassical limit.  The model is parameterized by the
vibrational normal mode frequencies, and the electronic energies and energy derivatives
at a reference configuration.  The approach has been used by our group to compute state-to-state
transition rates in semi-empirical models for organic semiconducting light-emitting diode and photovoltaics
\cite{tamura2008phonon,tamura2007exciton,bittner2014noise,singh2009fluorescence}.

We recently made a significant breakthrough in using this approach by tying it
it to a fully {\em ab initio} quantum chemical approach for determining the diabatic states
and electron/phonon coupling terms allowing unprecedented accuracy and utility
for computing state-to-state electronic transition rates.
Our methodology consists of two distinct components.
The first is the use of a diabatization schemes for determining
 donor and acceptor states in a molecular unit.
 The other is a projection scheme which enables us to analyze the contribution of vibrations in reactions. Similar decomposition schemes have been presented by Burghardt
  \cite{cederbaum2005short,gindensperger2006shortI,gindensperger2006shortII}
  and the approach used here builds upon the method given in Ref.~\citenum{pereverzev2009energy}.
  We recently benchmarked this approach against
both the experimental rates and
recent theoretical rates presented by Subotnik {\em et al.}
 \cite{subotnik2008constructing,subotnik2009initial,subotnik2010predicting}
 and successfully applied the approach to compute state-to-state transition rates
 in series of Pt bridged donor-acceptor systems recently studied by Weinstein's group.
We  review here these latter results along with the details of our methods.

\section{Theoretical Approach}\label{section:theory}
\subsection{Model Hamiltonian}
We consider a generic model for $n$ electronic states coupled linearly
to a phonon bath.  Taking the electronic ground state of the system as a reference
and assuming that the electronic states are coupled linearly to a common set of
 modes, we arrive at a generic form for the Hamiltonian, here written for two coupled
 electronic states:
 \begin{eqnarray}
H=\left(\begin{array}{cc}
\epsilon_{1} & 0\\
0 & \epsilon_{2}
\end{array}\right)+\left(\begin{array}{cc}
{\mathbf g}_{11}&{\mathbf g}_{12} \\
{\mathbf g}_{21} &{\mathbf g}_{22}
\end{array}\right)\cdot{\mathbf q} +\frac{{\mathbf p}^{2}}{2}+\frac{1}{2}\mathbf{q}^{T}\cdot\mathbf\Omega\cdot\mathbf{q}.
\nonumber \\
\label{ham1}
\end{eqnarray}
Here, the first term contains the electronic energies, $\epsilon_{1}$ and $\epsilon_{2}$ computed at a
reference geometry--typically that of the donor or acceptor state.   The second term represents the
linearized coupling between the electronic and nuclear degrees of freedom given in terms of the mass-weighted
normal coordinates $\mathbf q$. 
 The diagonal terms
give the adiabatic displacement forces between the reference geometry and the two states.  If we choose one of the
states as the reference state, then either $\mathbf g_{11}$ or $\mathbf g_{22}$ will vanish.
The remaining two terms correspond to the harmonic motions of the nuclear normal modes, given here in mass-weighted normal coordinates.
In the normal mode basis, the Hessian matrix, $\mathbf \Omega$,  is diagonal with elements corresponding to the normal mode frequencies, $\omega_{j}^{2}$.

We now separate Eq.~\ref{ham1} into diagonal and
off-diagonal terms
\begin{eqnarray}
\hat  H = \hat H_{o} + \hat V
\end{eqnarray}
and perform a polaron transform
using the unitary transformation~\cite{grover1970exciton,rice1994excitons,pereverzev2006time}.
\begin{eqnarray}
U&=&e^{-\sum_{ni}\!\!\frac{g_{nni}}{\hbar\omega_i}|n\rangle \langle
n|(a^{\dagger}_i-a_i)}
 \nonumber \\
&=&
\sum_{n}|n\rangle \langle n|e^{-\sum_{i}\!\!\frac{g_{nni}}{\hbar\omega_i}(a^{\dagger}_i-a_i)}
\label{unitary}
\end{eqnarray}
under which the transformed Hamiltonian is written in terms of the
diagonal elements
\begin{eqnarray} \tilde H_0=U^{-1}H_0U
=\sum_n\tilde\epsilon_n |n\rangle \langle
n|+\sum_i\hbar\omega_ia^{\dagger}_ia_i,
 \end{eqnarray}
with  the renormalized electronic energies,
\begin{eqnarray}
\tilde\epsilon_n=\epsilon_n-\sum_{i}\frac{g_{nni}^2}{\hbar\omega_i},
\end{eqnarray}
and off-diagonal terms,
\begin{eqnarray} \hat V_{nm}=\sum_{i}g_{nmi}\left(a^{\dagger}_i+
a_i-\frac{2g_{nni}}{\hbar\omega_i}\right)e^{\sum_{j}\frac{(g_{nnj}-g_{mmj})}{\hbar\omega_j}(a^{\dagger}_j-a_j)}.
\label{opm}
\end{eqnarray}
In the transformed (or dressed) picture the electronic transition from state
$|n\rangle$ to $|m\rangle$ is accompanied by the excitations of all the
normal modes.

At this point it is useful to connect the various terms in the phonon-dressed Hamiltonian
with specific physical parameters.
First, the reorganization energy is given by
\begin{eqnarray}
\lambda_{nm}=\sum_{j}\frac{\left(g_{nnj}-g_{mmj}\right)^{2}}{\omega_{j}} = \sum_{j}\hbar \omega_{j}S_{j}
\end{eqnarray}
where  the $\{S_{j}\}$  are  the Huang-Rhys factors for each phonon mode.
These are related to the Franck-Condon factor
describing the overlap between the $v_j=1$ vibronic state in one electronic state
with the $v_j=0$ vibronic state in the other.
Likewise, the energy difference between the renormalized energy gaps is related to the
driving force of the state-to-state transition,
\begin{eqnarray}
\Delta E_{nm} = \tilde \epsilon_n-\tilde \epsilon_m.
\end{eqnarray}
Transforming to the interaction representation
and performing a trace over the phonons gives the spectral density in
terms of the autocorrelation of the electron-phonon coupling
operators.
Using the explicit form of the electron-phonon coupling
operators, one can arrive at a compact expression for the
autocorrelation function of the electron/phonon coupling
\begin{widetext}
\begin{eqnarray}
C_{nm}(t) &=& \langle V_{nm}V_{mn}(\tau)\rangle\\
&=&
\sum_{i,j}g_{nmi}g_{mnj}
\left(\left(\Delta_{nmi}(\overline{n}_i+1)e^{i\omega_i\tau}
-\Delta_{nmi}\overline{n}_ie^{-i\omega_i\tau}+\Omega_{nmi}\right) \right.\nonumber \\
& &\times\left.\left(\Delta_{nmj}(\overline{n}_j+1)e^{i\omega_j\tau}
-\Delta_{nmj}\overline{n}_je^{-i\omega_j\tau}+\Omega_{nmj}\right)\right.\nonumber \\
& &\left.+\delta_{ij}(\overline{n}_i+1)e^{i\omega_i\tau}+\delta_{ij}\overline{n}_i
e^{-i\omega_i\tau}\right)q_{nm}(\tau)f_{nm}(\tau),
 \label{corrf}
\end{eqnarray}
\end{widetext}
Here, $\hat V_{nm}(t)$ is the electron-phonon coupling term in the Heisenberg representation and
$\langle \cdots \rangle$ denotes a thermal average over the
vibrational degrees of freedom.
The remaining terms are constructed from the normal mode frequencies $\{\omega_{i}\}$ and
electron/nuclear couplings $\{g_{nmi}\}$ viz.
\begin{eqnarray}
\Delta_{nmi}&=&\frac{(g_{nni}-g_{mmi})}{\omega_i}, \\
\Omega_{nmi}&=&\frac{(g_{nni}+g_{mmi})}{\omega_i}, \\
q_{nm}(\tau)&=&e^{i\sum_j{\Delta^2_{nmj}\sin\omega_j\tau}},\label{q} \\
f_{nm}(\tau)&=&e^{-2\sum_j(\overline{n}_j+\frac{1}{2})\Delta_{nmj}^2(1-\cos\omega_j\tau)}. \label{f}
\end{eqnarray}
Finally, $\overline{n}_{i}$ is the Bose population of vibrational normal mode $i$,
\begin{eqnarray}
\overline{n}_i=\frac{1}{e^{\beta\omega_i}-1}.
\end{eqnarray}
The spectral density and golden-rule rate can then be obtained by Fourier transform
\begin{eqnarray}
S_{nm}(\tilde\omega) = \int_{-\infty}^{\infty} \! \mathrm{d}t e^{-i\tilde \omega t}\langle \hat V_{nm}(t) \hat V_{mn}(0)\rangle.\label{spec-dens}
\end{eqnarray}
and
\begin{eqnarray}
k_{nm}=2{\rm Re}\int_{0}^{\infty} \! \mathrm{d}t\left\langle \hat V_{nm}(0)\hat V_{mn}\left(t\right)\right\rangle e^{-i\tilde\omega_{nm}t}.
\label{gr-expression}
\end{eqnarray}

\subsection{Non-Markovian Master Equation and Golden-Rule Rates}

The formalism presented above
requires both diagonal (${\mathbf g}_{nn}$) and off-diagonal  (${\mathbf g}_{nm}$) derivative couplings between
adiabatic states.   Until recently, these couplings were difficult to impossible to
compute for molecular systems using standard quantum chemical means.
We next discuss how we have used the Edmiston-Ruedenberg (ER) localization scheme \cite{edmiston1963localized} to estimate the couplings.
We also present how one can construct a reduced set of harmonic modes that
fully capture the electron/nuclear coupling.

\begin{figure}[!t]
\includegraphics[width=0.95\columnwidth]{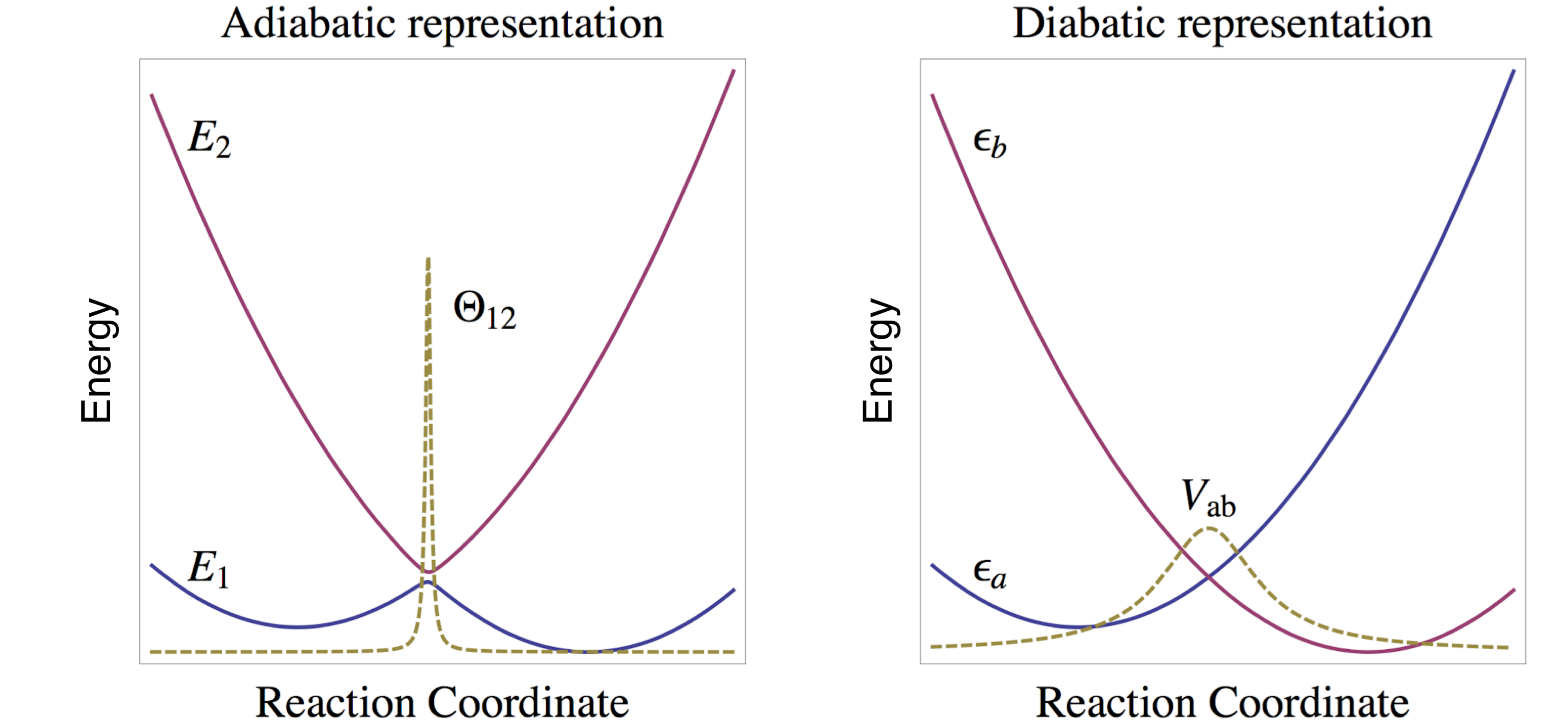}
\caption{Sketch of adiabatic and diabatic representations for two-state system.
 Compared to adiabatic representations, diabatic representation has smoother energy surfaces and couplings.
}
\label{adiaDiab}
\end{figure}

A workaround is to transform to a diabatic representation, whereby the Hamiltonian is written as
\begin{eqnarray}
H_{dia}&=&U^{T}H_{adia}U \nonumber \\
&=&\left(\begin{array}{cc}
\epsilon_{a}(\mathbf{R})\mathrm{+T'_{n}}(\mathbf{R})_{11} & V_{ab} \\
V_{ab} & \epsilon_{b}(\mathbf{R})\mathrm{+T'_{n}}(\mathbf{R})_{22}
\end{array}\right).
\end{eqnarray}
Fig.~\ref{adiaDiab} shows a sketch of the adiabatic and diabatic potentials for a model two level system.
While the adiabatic representation is precisely defined in terms of
electronic eigenstates, the diabatic representation offers several advantages.
First, the sharp derivative couplings that depend upon the nuclear velocity
 in the adiabatic representation are transformed to smoother  diabatic couplings, $V_{ab}$ that only depend upon the
 nuclear positions.
 Second, potential energy surfaces are  smoother and the avoided crossing is eliminated.
A number of diabatization approaches have been developed and the
reader is referred to  Ref.~\citenum{domcke2004conical} for a general review.

%----FIGURE 2
%-----
The problem now is how to obtain the transformation matrix
\begin{eqnarray}
U = \left(\begin{array}{cc}
\cos\theta & \sin\theta\\
-\sin\theta & \cos\theta
\end{array}\right).
\end{eqnarray}
While a number of methods are available %\cite{domcke2004conical},
%Here we introduce the Edmiston-Ruedenberg diabatization approach.
a straightforward approach is to eliminate derivative coupling mathematically by requireing
\begin{eqnarray}
\left\langle \phi_{i}(\mathbf{r};\mathbf{R})|\nabla_{\mathbf{R}}|\phi_{j}(\mathbf{r};\mathbf{R})\right\rangle = 0.
\end{eqnarray}
However, this is computationally very expensive--especially for complex molecular systems,
and exact solutions generally do not  exist. \cite{mead1982conditions}

An alternative approach is to use physical intuition rather than a purely mathematical constraint
to define the diabatic states.
The  Edmiston-Ruedenberg (ER)   diabatization scheme is such a scheme and is based on the idea that the diabatic
states can be obtained by maximizing the total electron repulsion between localized states,
\begin{eqnarray}
f_{ER}=\sum_{k}^{N_{states}}\int\int dr_{1}dr_{2}\frac{\left\langle \phi_{k}|\hat{\rho}\left(r_{1}\right)|\phi_{k}\right\rangle \left\langle \phi_{k}|\hat{\rho}\left(r_{2}\right)|\phi_{k}\right\rangle }{\left\Vert r_{1}-r_{2}\right\Vert }.
\end{eqnarray}
When the adiabatic (and diabatic) energy minima
are far enough away from the crossing points and the mixing angles between the diabatic  and adiabatic
states is small, we can
use the gradients of the adiabatic potentials to approximate the diabatic potentials.
Thus, if we perform calculations at the
optimized geometry of the final acceptor state  ({\em i.~e.} about $Q_{2}$  in Fig.~\ref{marcus}),
we can write the Hamiltonian as
\begin{eqnarray}
H_{dia,e}=\left(\begin{array}{cc}
\epsilon_{1} & V_{12}\\
V_{21} & \epsilon_{2}
\end{array}\right)+\left(\begin{array}{cc}
0 & 0\\
0 & 1
\end{array}\right) {\mathbf g}_{22}\cdot{\mathbf q}+H_{osc},
\end{eqnarray}
where $H_{osc}$ is the harmonic oscillator Hamiltonian for the vibrational normal modes.
The linear assumption amounts  to performing a series expansion of the
full, multi-dimensional coupling term and keeping only the lowest order terms.
Systematic improvement can be made by including higher-order (e.g.  quadratic) off-diagonal couplings.
However, this would involve a substantial increase in the complexity of the theory.
The linear assumption is reasonable so long as  the mixing angle is small.
\cite{xunmo1,xunmo2}

We obtain the diabatic couplings $V_{12}$
and the mixing angle $\theta$  via ER localization and transform the electronic Hamiltonian
from the adiabatic basis to the diabatic basis {\em viz.}
\begin{equation}
H_{dia}=\left(\begin{array}{cc}
\cos\theta & -\sin\theta\\
\sin\theta & \cos\theta
\end{array}\right)\left(\begin{array}{cc}
\epsilon_{1}  & 0 \\
0  & \epsilon_{2}
\end{array}\right)\left(\begin{array}{cc}
\cos\theta & \sin\theta\\
-\sin\theta & \cos\theta
\end{array}\right).\label{eq:boys}
\end{equation}
The diabatic coupling is then given by
\begin{eqnarray}
V_{ab}=\frac{1}{2}\sin2\theta\left(\epsilon_{2}-\epsilon_{1}\right).\label{mixingangle}
\end{eqnarray}

We then diagonalize the electronic part and transform the electron/nuclear coupling
back into the adiabatic basis.  In doing so,  we obtain the Hamiltonian in the
form given in Eq. \ref{ham1}
\begin{eqnarray}
H&=&U^{T}H_{dia}U\nonumber  \\
&=&\left(\begin{array}{cc}
E_{1}  & 0 \\
0 & E_{2}
\end{array}\right)+\left(\begin{array}{cc}
\sin^{2}\theta & \frac{1}{2}\sin2\theta\\
\frac{1}{2}\sin2\theta & \cos^{2}\theta
\end{array}\right) {\mathbf g}_{22}.{\mathbf q} \nonumber \\
&+& H_{osc}.\label{eq:locaiHam}
\end{eqnarray}

Alternatively, one can use the Generalized Mulliken-Hush model (GMH)  \cite{cave1996generalization,cave1997calculation}
which works well for linear systems but does not generalize easily to
systems with more than two charge centres.
Within GMH, the diabatic mixing is given by
$$V_{12}=\frac{\left(E_2-E_1\right) \left|\mu _{12}\right|}{\sqrt{\left(\mu _1-\mu _2\right){}^2+4 \mu _{12}^2}},$$
where $\left(E_2-E_1\right)$ is the vertical excitation energy, $\mu _1$ and $\mu _2$
are the dipole moments of corresponding adiabatic states,
and $\mu _{12}$ is the transition dipole moment between two states.
 ER localization can be seen as a extension of GMH which overcomes some drawbacks of GMH \cite{subotnik2009initial}.
Both ER and GMH require convergence of the initial and final reference states  and have be used
to compute the coupling terms required for the TCLME approach given above.\cite{xunmo1,xunmo2,xunmo3}

\subsection{Determining the Optimal Electron-Phonon Coupling Components}

While the Marcus expression is elegant in its simplicity in requiring three parameters that
can be obtained experimentally, it masks a wealth of detail that underlie the quantum transition.
Considerable insight into the state-to-state dynamics can be revealed by examining
the nuclear motions driving and coupling the electronic states.
Our approach is based on earlier work by our group
\cite{pereverzev2009energy} and Burghardt {\em et al.}
 \cite{gindensperger2006shortI,gindensperger2006shortII,cederbaum2005short}.
 Central to the theory is that there exists a collective nuclear displacement coordinate
that connects the initial geometry of the donor to the final geometry of the acceptor.
However, until this work a general systematic approach for determining such motions did not exist.

Generally speaking, this collective coordinate involves all nuclear degrees of freedom.
However, the form of the electronic Hamiltonian in Eq.~\ref{ham1} suggests that
there exists a subset of motions that are specific modes  that capture the majority of the
electronic/nuclear coupling and give a dominant contribution to the collective
reaction coordinate.  Within the linearized approximation for the electronic/nuclear coupling,
we can write a force tensor
\begin{eqnarray}
{\mathbf F} =
\left(\begin{array}{cc}
{\mathbf g}_{11}&{\mathbf g}_{12} \\
{\mathbf g}_{21} &{\mathbf g}_{22}
\end{array}\right)
\end{eqnarray}
where ${\mathbf F}\cdot \mathbf q$ is the electronic/nuclear coupling term in Eq.~\ref{ham1}.
If we consider each unique element $\{ \mathbf g_{11}, \mathbf g_{12} , \mathbf g_{22}\}$ to be
linearly independent, but non-orthogonal force vectors,  one can develop a  projection operator scheme to
to parse the $N$-dimensional linear vector space spanned by the mass-weighted normal mode vectors into two subspaces:
one spanned by three vectors describing the coupling between the electronic states
and the other spanned by the remaining $N-3$ dimensional space spanned by motions that
do not couple the electronic states.
This  subspace  can be generated by defining a projection operator
\begin{eqnarray}
\mathbf{P}=\sum_{\alpha\beta}'\left(\mathbf{S^{-1}}\right)_{\alpha\beta}\mathbf{g_{\alpha}}\otimes\mathbf{g_{\beta}}
\end{eqnarray}
in which the summation is limited to linearly independent vectors.
   Here $\mathbf{S}_{\alpha\beta}=\mathbf{g_{\alpha}}\cdot\mathbf{g_{\beta}}$,
$\otimes$ is outer product,  and $\mathbf{I}$ is unitary operator.
This $N\times N$ matrix projects out all normal modes that are directly coupled to the
electronic degrees of freedom and
 its complement $\mathbf{Q}=\mathbf{I}-\mathbf{P}$ projects out all modes not directly coupled.
By diagonalizing the matrix
\begin{eqnarray}
\mathbf{K}=\mathbf{P}\cdot\mathbf\Omega\cdot \mathbf{P}+\mathbf{Q}\cdot\mathbf\Omega\cdot\mathbf{Q}
\end{eqnarray}
we obtain a  transformation, ${\mathbf M}$,  between the normal coordinates and a new set of orthogonal
coordinates.  Both $\mathbf{P}\cdot\mathbf\Omega\cdot \mathbf{P}$ and $\mathbf{Q}\cdot\mathbf\Omega\cdot\mathbf{Q}$ are $N\times N$ matrices.
However, for a two-state system, the former will have exactly $3$ non-trivial eigenvalues, $\{\alpha_{p}\}$, with corresponding
eigenvectors, $\{ M_{p}\}$, whereas the latter will have exactly $N_{r} = N-3$ non-trivial eigenvalues, $\{\alpha_{q}\}$,  and corresponding
eigenvectors, $\{M_q\}$.   This the full $N\times N$ transformation is formed by joining the non-trivial vectors from the
two respective subspaces ${\mathbf M} = \{M_{p}, M_{q}\}$.
The transformed electron-phonon coupling constants are given by projecting the  couplings   in the normal mode basis on to the new
basis.
\begin{eqnarray}
\mathbf{g}_{ab}'=\mathbf{M}_{p}\cdot\mathbf{g}_{ab}.
\end{eqnarray}
By examining the types of molecular motions that compose the ${\mathbf M_{p}}$ subspace, we can
gain a deeper understanding of the specific classes of internal motion that are directly involved with the
electron transfer process.  In addition,  we can gain a computational advantage since presumably this
 reduced set of modes give the dominant contribution to the electron-phonon coupling and  autocorrelation
 function given as the kernel in Eq. ~\ref{gr-expression}.

It is crucial to notice that the vectors  given in Eq.~\ref{eq:locaiHam} are {\em not linearly independent}.
 Consequently, special care
must be taken to generate the reduced sub-space.  To facilitate this, we develop an iterative Lanczos approach,
taking the normalized vector ${\mathbf v}_{1} = {\mathbf g}_{22}$ as a starting point.
As above, we initialize each step indexed by $k$,  by defining a projection operator
\begin{eqnarray}
{\mathbf P}_{k} = {\mathbf v}_{k}\otimes  {\mathbf v}_{k}
\end{eqnarray}
and its complement ${\mathbf Q}_{k} = {\mathbf I}  - {\mathbf P}_{k}$.
%${\mathbf P}_{k}$ is the projection operator
for the $k$-th mode.
We also construct
\begin{eqnarray}
\mathbf p = \sum_{k} \mathbf P_{k}
\end{eqnarray}
as  the total projection operator for all ${k} \le N$ modes.
We then project the Hessian matrix $\mathbf\Omega$ into each subspace {\em viz.}
\begin{eqnarray}
\mathbf\Omega_{p} = \mathbf P_{k}\cdot \mathbf\Omega \cdot \mathbf P_{k} \,\, \& \,\, \mathbf\Omega_{q} = \mathbf Q_{k}\cdot \mathbf\Omega \cdot  \mathbf Q_{k}
\end{eqnarray}
and diagonalize each to obtain eigenvalues and eigenvectors $\{\alpha_{p}, {\mathbf M}_{p}\}$ and $\{\alpha_{q}, {\mathbf M}_{q}\}$
respectively.
As above, $\mathbf\Omega_{p} $ and $\mathbf\Omega_{q}$ are $N\times N$ matrices.
The first set will have a single
non-trivial eigenvalue and the second set
will have $N-k$ non-trivial eigenvalues.  As above we collect the non-trivial eigenvectors associated with each
to form the orthogonal transformation matrix
\begin{eqnarray}
{\mathbf M}_{k} = \{{\mathbf M}_{p},{\mathbf M}_{q}\},
\end{eqnarray}
and again transform the full Hessian $\mathbf\Omega$ into this new vector space to form the $N\times N$ matrix $\mathbf\Omega'$.
 At each step in the iteration, the transformed Hessian, $\mathbf\Omega'$ is in the form of a
$k\times k$ tri-diagonal submatrix in the upper-left part of the matrix and
a diagonal submatrix in the lower-right.  For example, after $k=3$ iterations, the Hessian matrix takes the form:
\begin{eqnarray}
{\mathbf\Omega}' =
\left[
\begin{array}{ccccccc}
\alpha_{1}   & b_{1}    & 0                    &               &             &            & 0 \\
b_{1}     & \alpha_{2}  & b_{2}                         \\
0            &   b_{2}         & \alpha_{3}    &   c_{k+1}   & c_{k+2}  & \cdots & c_{N}     \\
               &                     &           c_{k+1}           &     \alpha_{k+ 1} & &             & 0                 \\
               &                     &           c_{k+2}           &                 & \alpha_{k+2}  \\
               &                     &          \vdots           &                 &                    &  \ddots \\
   0           &                     &           c_{N}           &     0   &&& \alpha_{N}\\
\end{array}
\right]
\label{omega-prime}
\end{eqnarray}
We note that only the $k$-th mode is coupled the $N-k$ remaining modes.
Since all of the transformations are orthogonal, diagonalizing $\mathbf\Omega'$ at any point
returns the original Hessian matrix.

To continue iterating, we  take the $k$-th row of $\mathbf\Omega'$ and zero the first $k$ elements
$$
{\mathbf e} = \{0,\cdots 0,c_{k+1},c_{k+2},\cdots ,c_{N}\}.
$$
This is the coupling between the upper tridiagonal block and the lower diagonal block.
We thus obtain a new vector
$$
{\mathbf v}_{k+1} = {\mathbf e} \cdot {\mathbf M}
$$
which is then reintroduced into the iteration scheme.

For the first iteration, ${\mathbf v}_{1}$ is parallel to the bare electron-phonon coupling vector $g_{22}$
and the associated frequency is ${\mathbf v}_{1}\cdot\Omega\cdot{\mathbf v}_{1}$.   The subsequent iterations introduce
corrections to this via phonon-phonon coupling mediated via the electronic couplings.  For example, for the $k=3$ iteration,
we would determine the active vector space in terms of the upper-left 3$\times 3$ block of the matrix in
Eq.~\ref{omega-prime}.
\begin{eqnarray}
{\mathbf \Omega}_{3}' =
\left[
\begin{array}{ccc}
\alpha_{1}   & b_{1}    & 0      \\
b_{1}     & \alpha_{2}  & b_{2}                         \\
0            &   b_{2}         & \alpha_{3}
\end{array}
\right]
\end{eqnarray}
Diagonalizing  ${\mathbf \Omega}_{3}'$ returns a set of frequencies and  associated eigenvectors
which are then used to compute the electron-phonon couplings in this reduced active space.
After $N-1$ iterations, $\mathbf\Omega'$ is a fully tridiagonal matrix and diagonalizing this returns the original
normal mode basis.

At any point along the way, we can terminate the iteration and obtain a reduced set of
couplings.  Since the  Lanczos approach uses the power method for finding the largest eigenvector of a matrix,
it converges first upon the vector with the largest electron/nuclear coupling--which
 we refer to as the ``primary mode''.  Subsequent iterations produce reduced modes with progressively
 weaker electron/nuclear couplings and the entire  process can be terminated after a few iterations.
After $k$-steps, the final electron-phonon couplings are then obtained by
projecting the original set of couplings (in the normal mode basis) into the final vector space.
For small systems, we find that accurate rates can be obtained with as few as 2 - 3 modes are
sufficient to converge the autocorrelation function in Eq.~\ref{corrf}.\cite{xunmo2,xunmo1}

\section{Inelastic electronic coupling in Donor-Bridge-Acceptor complexes}\label{chap:chapt4}

\begin{figure*}[t]
\subfigure[]{\includegraphics[width=\columnwidth]{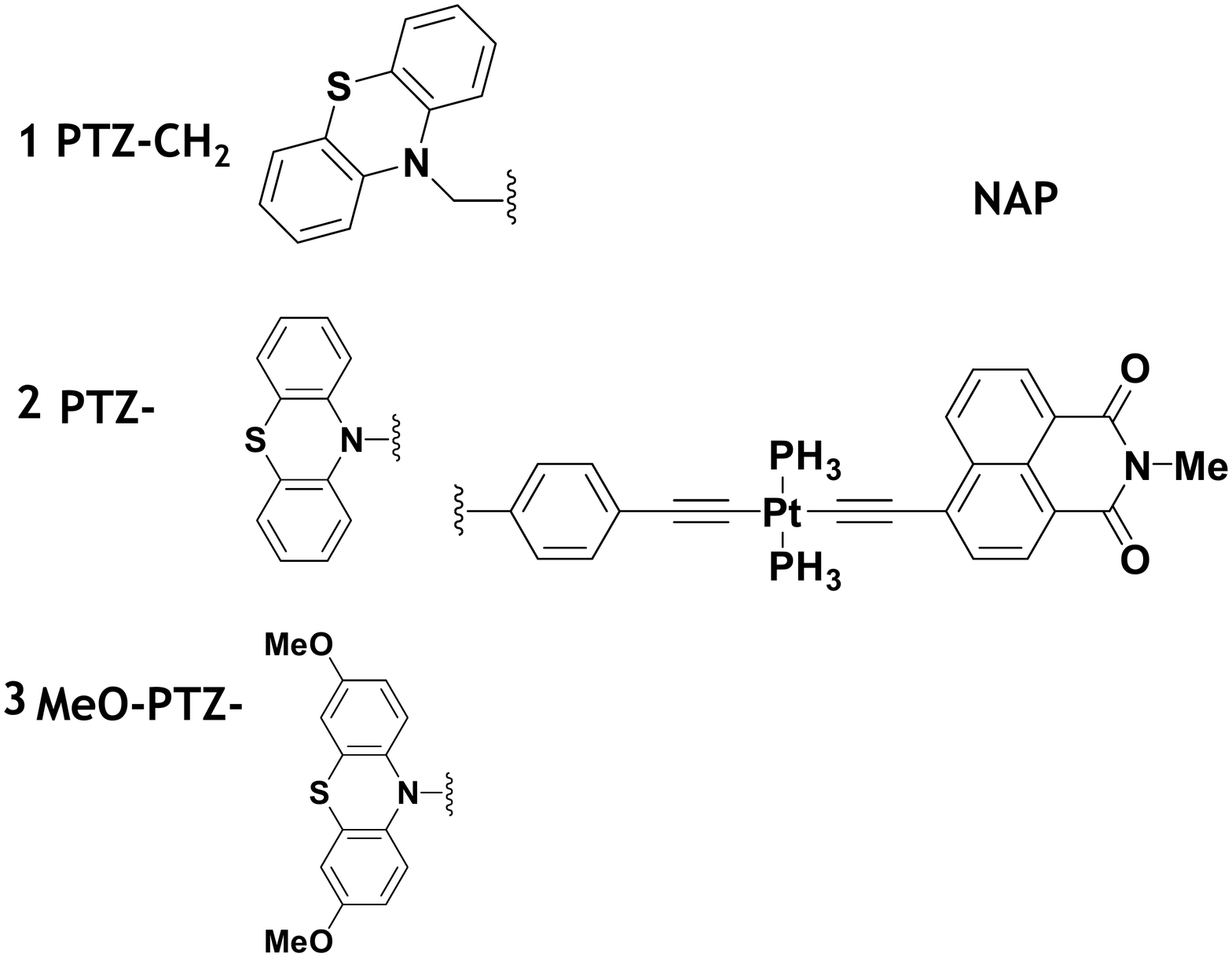}}
\subfigure[]{\includegraphics[width=\columnwidth]{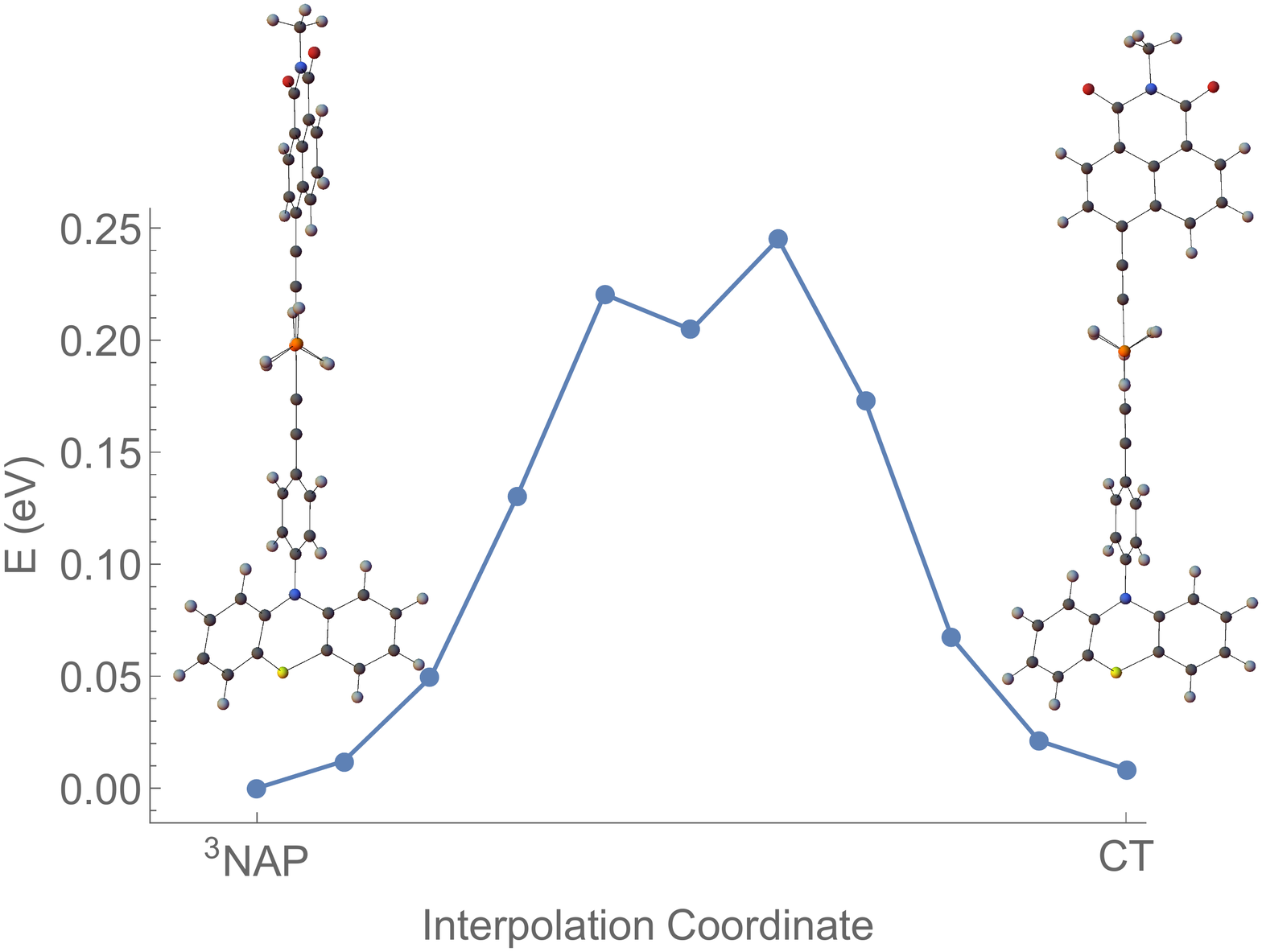}}
\caption{(a.) Chemical structures of the Donor (P), Bridge (-Pt-), Acceptor (NAP) complexes
considered here.  (b.) Triplet energy along a linear interpolation coordinate
connecting the $^{3}$NAP minimum energy geometry and the CT  minimum energy geometry.
\label{Fig2}}
\end{figure*}

The Weinstein group at University of Sheffield reported recently upon a series of
 donor-bridge-acceptor (DBA) molecular triads
 whose electron transfer (ET) pathways can be radically changed
 - even completely closed - by infrared light excitation of specific
 intramolecular vibrations \cite{delor2014toward,delor2015mechanism,scattergood2014electron}.
 The triads consist of a phenothiazine-based (PTZ) donor linked to a naphthalene-monoimide (NAP) acceptor via a Pt-acetylide bridging unit.\cite{scattergood2014electron} The structures of the triads are given in~Fig. \ref{Fig2}a.  All three systems undergo a similar sequence of processes after following UV excitation: electron transfer from the Pt-acetylide center to the NAP acceptor, resulting in a charge-transfer state, $D-B^+-A^-$, which due to strong spin-orbit coupling efficiently populates triplet charge-transfer state, CT. Further electron transfer leads to a fully charge-separated state (CSS) $D^+-B-A^-$ with the electron and hole localized on the acceptor and donor units respectively. The charge transfer state can also undergo charge recombination to form a localized triplet exciton on the NAP unit ($^3$NAP), or the ground state. Both CSS and $^3$NAP decay to the singlet ground state on the nanoseconds and sub-millisecond time scales, respectively.
 We also show in Fig.~\ref{Fig2}b the triplet energy along a linear interpolation coordinate connecting the $^3$NAP minimum energy geometry to the
 CT minimum energy geometry.  Between the two is a significant energy barrier reflecting the
relative rotation of the NAP and the PTZ groups about the CC-Pt-CC axis.

%-----FIGURE 3

The UV pump-IR push experiments performed on these triads showed that IR-excitation of bridge vibrations after the initial UV pump radically changes
the relative yields of the intermediate states.
Subsequent excitation of the -CC-Pt-CC- localized vibrations by a timed IR pulse
in the CT state of PTZ-complex {\bf 2} at 1 ps after the UV pump decreases the yield of the CSS state,
whilst increasing that of the $^3$NAP state.
IR-excitation in the course of electron transfer has caused a 100\% decrease in the CSS yield in {\bf 1},
approximately 50\% effect in {\bf 2}, and no effect in {\bf 3}.

This demonstration of control over excited state dynamics strongly suggests that the acetylide stretching modes are significantly involved in the electron/nuclear coupling in these systems and play central roles in the electron-transfer process.
 The transferred charge can undergo either further separation to form the full charge-separated state (CSS), or recombine to form a localized excitation, $^3$NAP. Both eventually decay to ground state. Weinstein {\em et al.} showed that if a
 judiciously chosen IR pump is applied to excite the C$\equiv$C bond after the initial UV excitation, the yield of intermediate states can be radically changed.
For example, when a IR pump with frequency = 1,940 $cm^{-1}$ is applied to excite the C$\equiv$C in PTZ-CH$_{2}$-Pt-NAP, 1 ps after the UV pump, the yield of the electron transfer
   state decrease from 32\% to 15\%, while that of the $^3$NAP increases from 29\% to 46\%. The most striking observation is that when a 1,908 $cm^{-1}$ IR pulse is applied to PTZ-CH$_{2}$-Pt-NAP 2 ps after UV excitation, the CT $\rightarrow$
   CSS step is completely switched off.
 \cite{delor2014toward,delor2015mechanism,scattergood2014electron}.

Quantum chemical analysis indicates that the electron-transfer rate is largely influenced by chemical modification of the PTZ donor.
From PTZ-CH$_2$-Pt-NAP to PTZ-Pt-NAP, to OMe-PTZ-Pt-NAP, the donor strength increases, which
increases the energy gap between CT and CSS states.
  The driving force ($\Delta G$) for the CT $\rightarrow$ CSS transfer
  also increases from 0.2 eV in PTZ-CH$_2$-Pt-NAP, to 0.4 eV in PTZ-Pt-NAP, to 0.6 eV in
  OMe-PTZ-Pt-NAP. Large $\Delta G$ accelerates the CT decay and hence decreases the lifetime of CT state.
Comparing  PTZ-Pt-NAP and PTZ-CH$_2$-Pt-NAP, CT transfer to both charge separation and recombination
 slows down by a factor of  about 5 (the lifetime of CT increases from 3.3 to 14 ps and CSS from 190 ps to 1 ns).
By appending methoxy groups to the PTZ, the donor strength is increased, and reaction is accelerated.
As the result, the lifetime of CT in OMe-PTZ-Pt-NAP is further reduced to 1 ps.
Weinstein {\em et al.} proposed that
 the effect of infrared control is caused by the fact that the distance between CT energy minimum and the intersection of CT and CSS potential energy surfaces is small. For all three molecules, two PESs intersect where C$\equiv$C bond is slightly longer than the equilibrium length. When C$\equiv$C bond gets excited, it elongates and helps molecules to pass the intersection. If the energy gap between intersection and equilibrium geometry is much larger than C$\equiv$C vibrational energy, the dynamics is barely affected; if the energy gap is small, the vibrational excitation can radically change the dynamics.
\cite{delor2014toward,delor2015mechanism,scattergood2014electron}.

\subsection{Theoretical Model}

\begin{figure}[b]
{\includegraphics[width=\columnwidth]{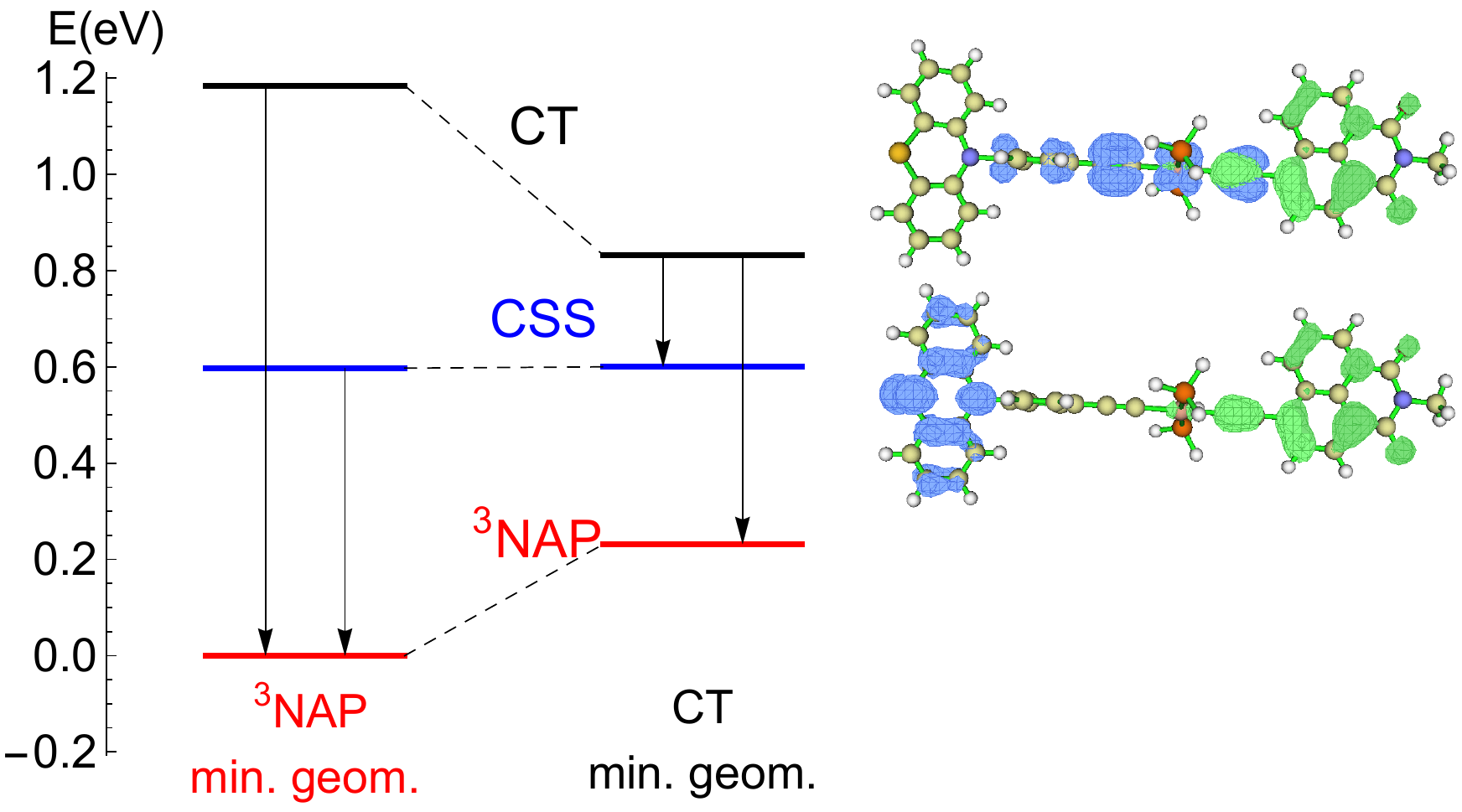}}
\caption{ Energy level diagram for the  triplet states of PTZ at the $^3$NAP and CT state geometries.
The electron/hole distributions for the CT and CSS are shown to the right (green=electron, blue=hole).
}
\label{ene_diagram}
\end{figure}

We focus our attention on the PTZ system and anticipate that the
 other systems in this study will exhibit similar
 behaviour due to the overall similarity of the
various donor groups.\cite{xunmo3}
 For purposes of facilitating the calculations,
 the molecular structures are simplified such that the P(Bu)$_{3}$ moieties and octyl chain of the NAP group were
 truncated to -PH$_{3}$ and a single methyl group, respectively. In all quantum chemical calculations,
 we used the SDD pseudo-potential for Pt and 6-31G(d, p) for the other atoms.
 We also used the Polarizable Continuum Model (PCM)  to account for the
 dichloromethane solvent%\cite{mennucci1997continuum,cossi1998ab}
as used in Ref.\citenum{delor2014toward,delor2015mechanism,scattergood2014electron}.
The transition dipole moments and electron/hole distributions surfaces were calculated using the  Multiwfn (v3.3.8)  program.\cite{lu2012multiwfn} 
An energy level diagram based upon our calculations is sketched in Fig. \ref{ene_diagram}a
together
with the corresponding electron/hole distribution plots.

%-------TABLE 1
\begin{table*}[t]
\centering

%\begin{adjustbox}{max width=\columnwidth}

\begin{tabular}{c|c|c|c|c|}
             & \multicolumn{2}{|c|}{$^3$NAP Geom. (0~eV)}             & \multicolumn{2}{c|}{CT Geom. (0.818~eV)}           \\
             & CSS $\rightarrow$ $^3$NAP & CT $\rightarrow$ $^3$NAP & CT $\rightarrow$ $^3$NAP & CT$
\rightarrow$CSS \\
\hline
$\Delta G^{\circ}$ (eV)           & 0.414                & -0.913              & -0.781              & -0.20           \\
$\lambda$ (eV)             & 1.01                 & 0.271               & 1.38                & 1.08               \\
$V$(eV)                     & 2.56E-4              & 0.345               & 1.34E-2             & 9.22E-3            \\
$\overline{V}$ (eV)                 & 6.34E-2              & 0.106               & 0.106               & 0.192              \\
$\Delta G^{\circ}_{\bar{V}}$ (eV) & 0.414                & -0.851              & -0.770              & N/A\\

\end{tabular}

%\end{adjustbox}
\caption{Driving force $\Delta G^{\circ}$, reorganization energy $\lambda$, diabatic coupling $V$, mean diabatic coupling $\overline{V}$, and $\Delta G^{\circ}_{\bar{V}}$ (driving force calculated with $\overline{V}$), for different transitions}
\label{MarcusParam}
\end{table*}

To obtain the diabatic potentials and couplings, we perform a geometry optimization of both the lowest triplet ($^3$NAP) and the third triplet  excited states (CT).
As discussed below, we use the optimized states as reference geometries for determining the diabatic coupling within the
Generalized Mulliken Hush approximation.\cite{cave1996generalization,cave1997calculation}
The normal modes and vibrational frequencies were obtained by harmonic expansion of the energy
about the CT state.
Once we have determined the diabatic states and couplings,
we use the TCLME approach from Ref.~\cite{pereverzev2006time} to compute the time-correlation functions  and
 state-to-state golden-rule rates as discussed above.
 We also use the projection technique to determine an optimal set of normal modes and determine
the number of  such optimal modes that are required to converge the time-correlation functions to a desired degree of accuracy.
We then use both the CT and  $^3$NAP minima as reference states for computing
the diabatic potentials and couplings necessary for computing rates and modes.
Those obtained at the CT minimum can be used to compute transitions originating in from the CT state,
while those obtained at the  $^3$NAP minimum can be used for transitions terminating
in the $^3$NAP state.

%-------FIGURE 4

%------FIGURE 5.
\begin{figure}[t]
\subfigure[]{\includegraphics[width=0.41\columnwidth]{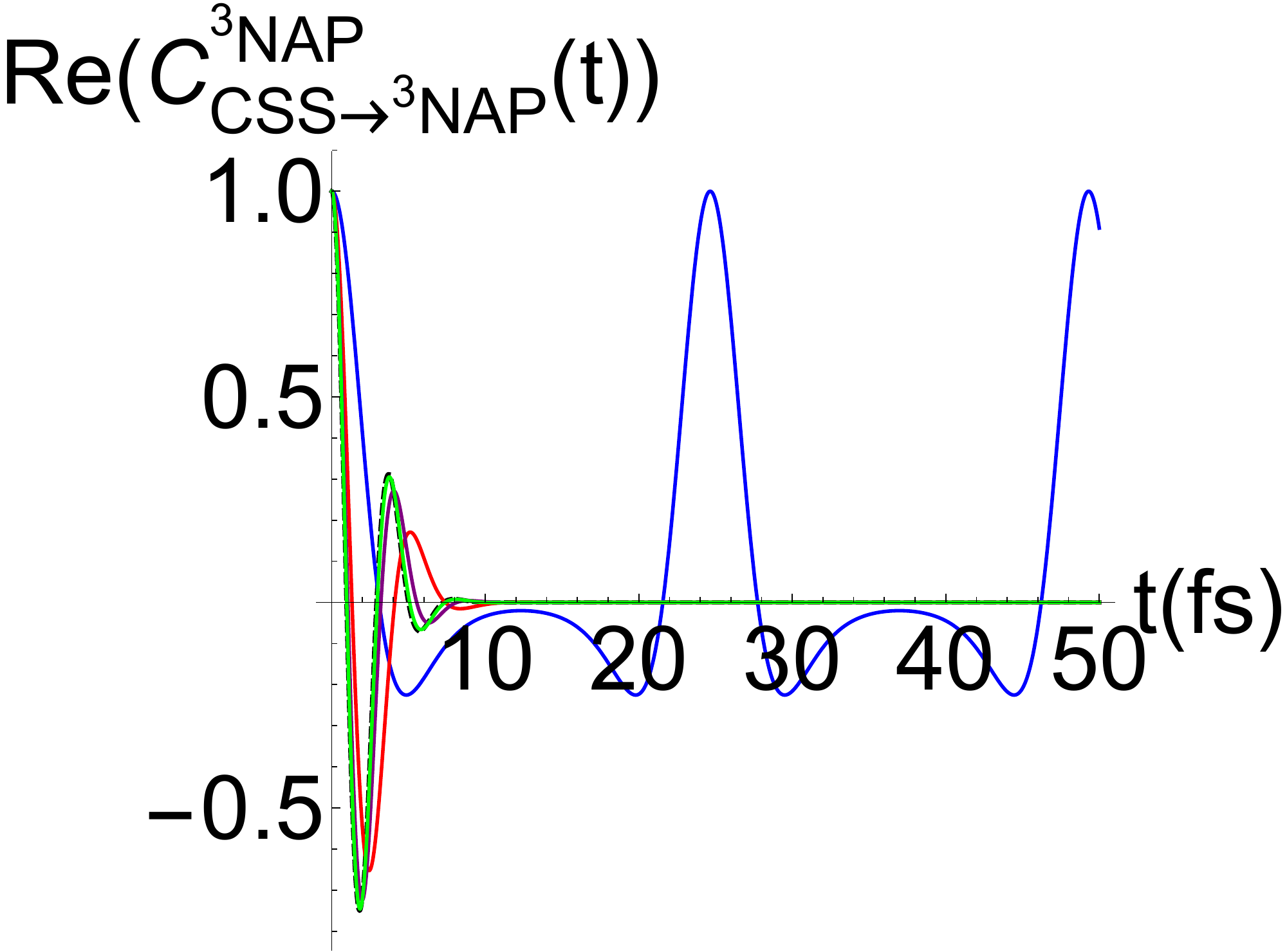}}
\subfigure[]{\includegraphics[width=0.57\columnwidth]{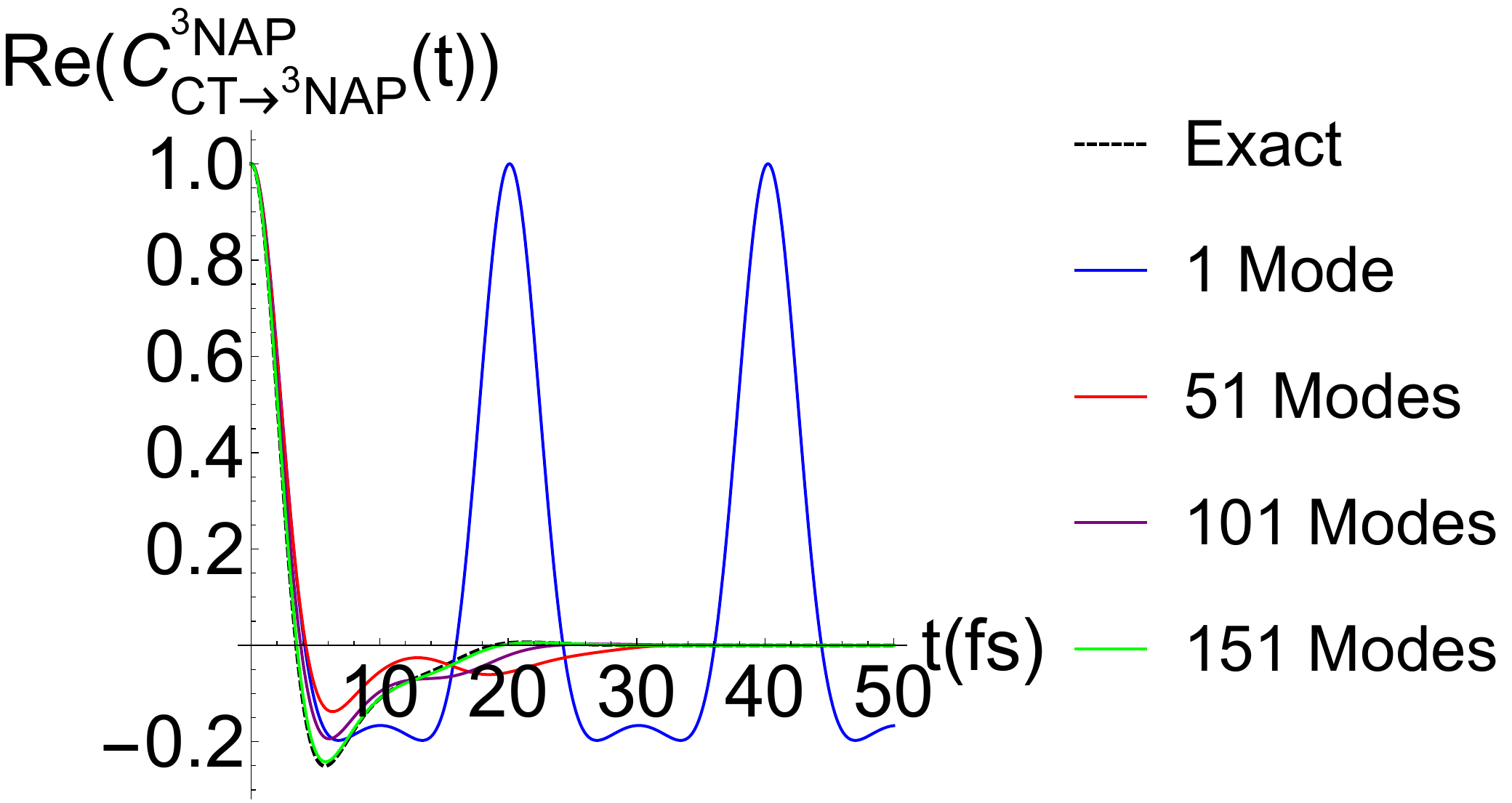}}\\
\subfigure[]{\includegraphics[width=0.41\columnwidth]{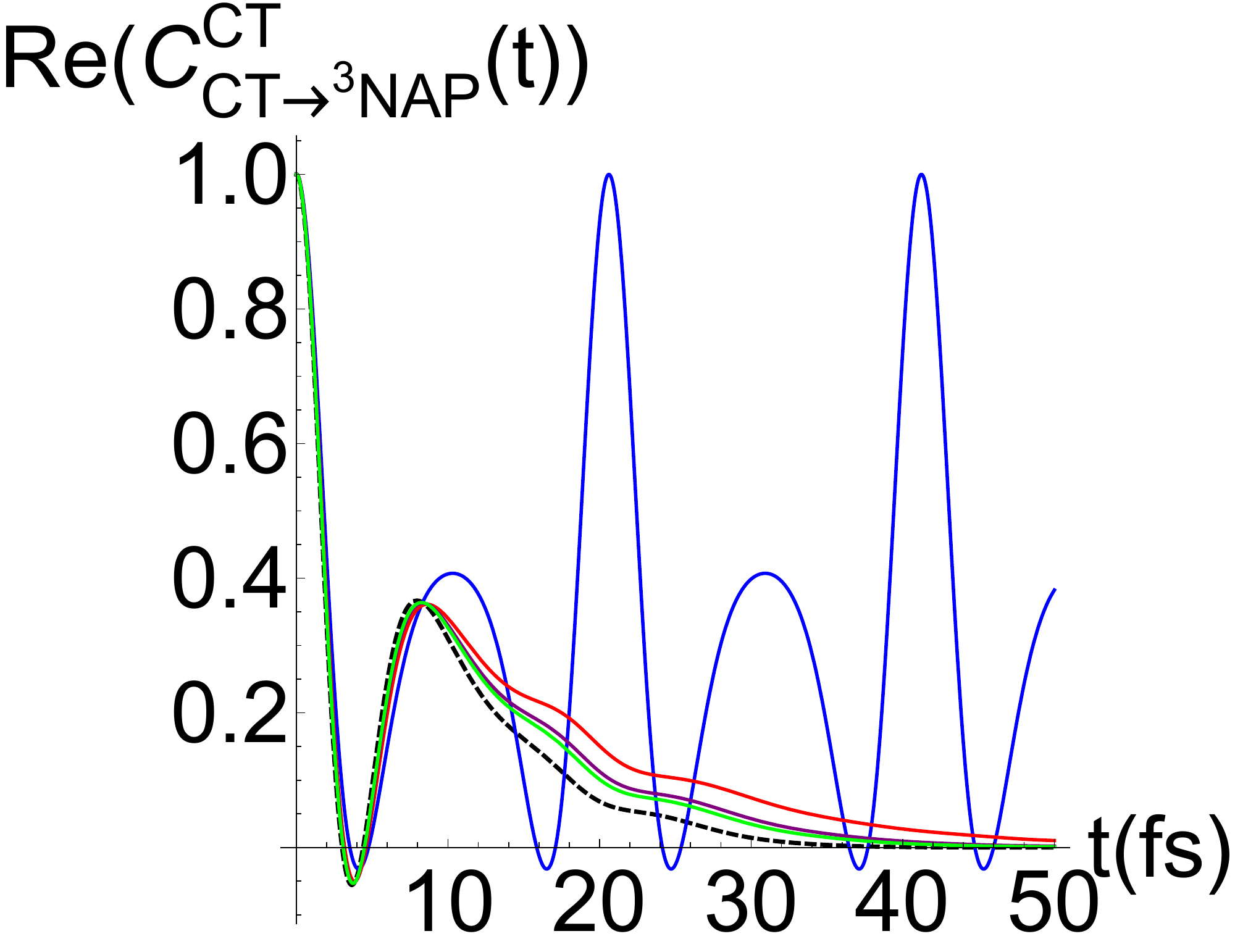}}
\subfigure[]{\includegraphics[width=0.57\columnwidth]{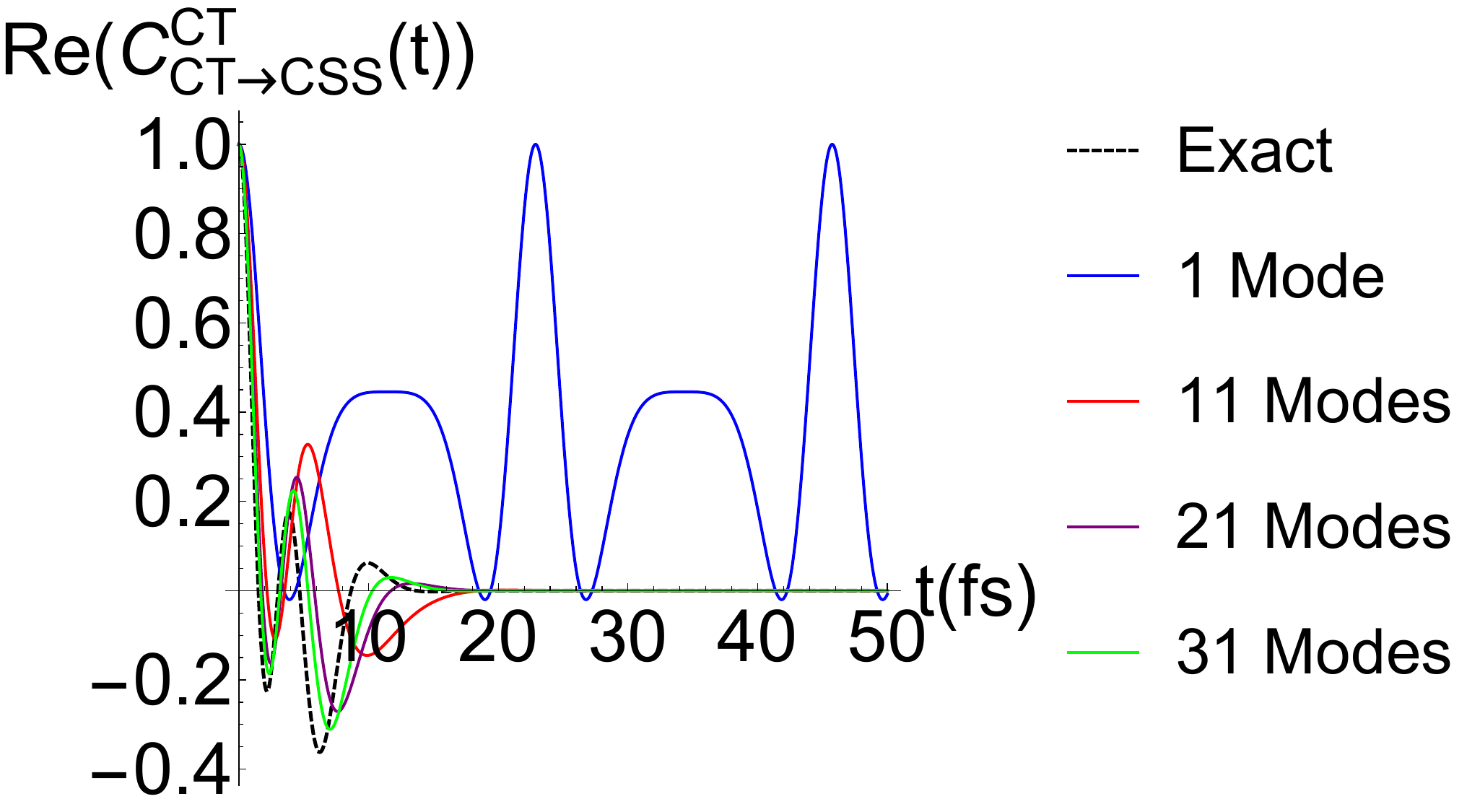}}
\caption{
Correlation functions of various numbers of projected modes, compared to exact correlation, for (a) CSS $\rightarrow$ $^3$NAP at $^3$NAP geometry, (b) CT $\rightarrow$ $^3$NAP at $^3$NAP geometry, (c) CT $\rightarrow$ $^3$NAP at CT geometry, and (d) CT $\rightarrow$ CSS at CT geometry.
}\label{corrT1T3}
\end{figure}

We now compare electron transfer rates as computed using both Marcus theory
and the TCLME approach.  In the latter case, we examine the convergence of both the
time-correlation functions and the rate constants with respect to the number of
nuclear modes included in the summation in the construction of the electron-phonon coupling in Eq.~\ref{opm}.
For our purposes,  an ``exact''  calculation involves including all nuclear vibrational modes.
In our previous work we showed that both $C(t)$
and the total transfer rate constant, $k_{nm}$ calculated using only the first few projected modes provide an
excellent agreement with the exact quantities computed using  the full set of
normal modes, as well as the experimental rates,
when parameterized using accurate quantum chemical data\cite{xunmo1,xunmo2}.
\subsection{Marcus theory rates}
The Marcus expression  provides a succinct means for the computing 
transition rates from the driving force $\Delta G^{\circ}$,  diabatic coupling $V_{ab}$ and 
reorganization energy $\lambda$ in Eq.~\ref{marcus}.
In Table~\ref{MarcusParam}, we provide a summary 
of the parameters computed for the transitions we are considering . 
The two columns under the heading labeled $^{3}$NAP correspond to parameters computed using the $^{3}$NAP minimum as a
reference geometry while those under the heading labeled
 CT correspond to parameters computed using the CT reference
geometry.    
The Marcus rates provide a useful benchmark for our approach. 
Moreover, the parameters in this table portend a 
difficulty in using the $^{3}$NAP geometry as a reference. 
For example, for the CSS $\rightarrow^3$NAP transition, the driving force is in the wrong 
direction since it predicts that the CSS state lies lower in energy than the $^{3}$NAP state,
which is inconsistent with both experimental observations and our quantum chemical 
analysis in Figure \ref{ene_diagram}.

\begin{table*}
\begin{tabular}{l|r|r|r|r|}
                  &\multicolumn{2}{|c|}{CT Geom.}&\multicolumn{2}{|c|}{$^3$NAP Geom.}\\
 Rates (ps$^{-1}$)&  CT$\to$ CSS  &  CT $\to^3$NAP  &  CT $\to^3$NAP  &  CSS$\to^3$NAP  \\
\hline
 Exp.               &       0.0879  &         0.097  &           0.097  &        1.84E-3  \\
 Marcus             &        0.846  &        0.2043  &         1002.82  &      8.250E-11  \\
 Marcus (Mean V)    &        365.7  &         12.75  &           95.23  &        5.04E-6  \\
 TCLME              &        0.725  &        0.0562  &           12.89  &       3.022E-8  \\
 TCLME + PLM        &        0.627  &        0.0488  &            21.6  &       0.500E-4  \\
 TCLME (Mean V)     &           --  &          2.79  &           8.931  &        1.51E-3  \\
\end{tabular}
\caption{Comparison between experimental and computed state to state transition rates for PTZ.
The experimental rates for each process are obtained from Ref. \citenum{delor2015mechanism}.
}
\label{RateSummary}
\end{table*}
\subsection{TCLME Rates}
To compute the  rates using the TCLME expression (Eq. \ref{gr-expression}),
we begin by computing the
electron/nuclear correlation function and compare its convergence with respect to the
number of Lanczos modes.
Recall that the Lanczos modes are determined by an iterative ranking algorithm that identifies superpositions
of normal modes that optimize the electron/phonon coupling.
Fig.~\ref{corrT1T3}  gives a summary of these numerical tests in which we compute $C_{nm}(t)$
vs. time with an increasing number of Lanczos modes.     In all cases, we compare to the ``exact'' result in which
all nuclear modes were used.  The top two figures (Fig.~\ref{corrT1T3}a and b) use the $^{3}$NAP as the reference
geometry.  In these cases, convergence  of $C_{nm}(t)$ with respect to the number of modes proved to
be problematic for both transitions considered.    Correspondingly, the rates computed
using this geometry also compare poorly against the observed experimental rates, although are
an order of magnitude closer than Marcus rates.
We speculate that this may signal a break-down in the Condon approximation
which insures separability between nuclear and electronic degrees of freedom.

In Table~\ref{RateSummary}, we summarize both the experimental and computed state to state rates  for the PTZ
system.  Given the complexity and size of the system, overall the numerical rates computed using the
exact TCLME approach are in quantitative agreement with the experimental rates, particularly for those
using the CT geometry as a reference point (cf Fig.~\ref{corrT1T3}c and d).
We note that  fewer projected modes (30-50) are
needed to converge the correlation function out to the first 50~fs when using the CT-geometry.
Furthermore, while Marcus rate for the CT$\to$ CSS transition agrees with the exact TCLME
result, it misses the CT $\to^3$NAP experimental rate by 4 orders of magnitude whereas the TCLME rate
is in much better agreement with the experimental rate.

If we compare the exact TCLME rate, which uses the full set of normal modes in constructing the $C_{nm}(t)$
correlation function, to the rate computed used {\em only} the PLM (TCLME+PLM),
for both the CT$\to$ CSS  and CT $\to^3$NAP rates, the single mode approximation
is  within 86\% of the exact result.  This indicates that while multiple vibrational normal modes
contribute to the electronic coupling, the linear combination identified by the projection algorithm
carries the vast majority of the electron/phonon coupling.   This is  consistent with our
previous study of triplet energy transfer in small donor-bridge-acceptor systems.
\cite{xunmo1,xunmo2}

\subsection{Primary Mode Approximation}

As discussed earlier, our ranking algorithm allows us to rapidly determine the vibrational motions that optimize the
electron/nuclear couplings.  In addition to providing an accurate way to compute rate constants, they provide additional
insight into actual dynamics.  Here, we shall focus upon the transitions originating from the CT geometry.
Generally speaking, the highest ranked mode, termed the ``Primary Lanczos Mode'' (PLM),
 captures much of the short-time dynamics of the transitions.
In Fig. \ref{corrT1T3}, we show the electronic coupling correlation functions computed using
 different numbers of projected modes for all four transitions.
 For the CT $\rightarrow$ $^3$NAP transition, the
primary mode resembles the exact initial dynamics for the first 10  fs and
roughly 10 or so modes are sufficient to converge the correlation function out to times longer than the
correlation time.
 In Table \ref{RateSummary} we see that for the CT geometry, the primary mode approximation
is sufficient to obtain accurate rate constants.
 On the contrary, it takes considerably more modes modes to recover the full correlation function
 for transition originating from the $^3$NAP geometry.

\begin{figure*}[]
\subfigure[]{\includegraphics[width=0.45\columnwidth]{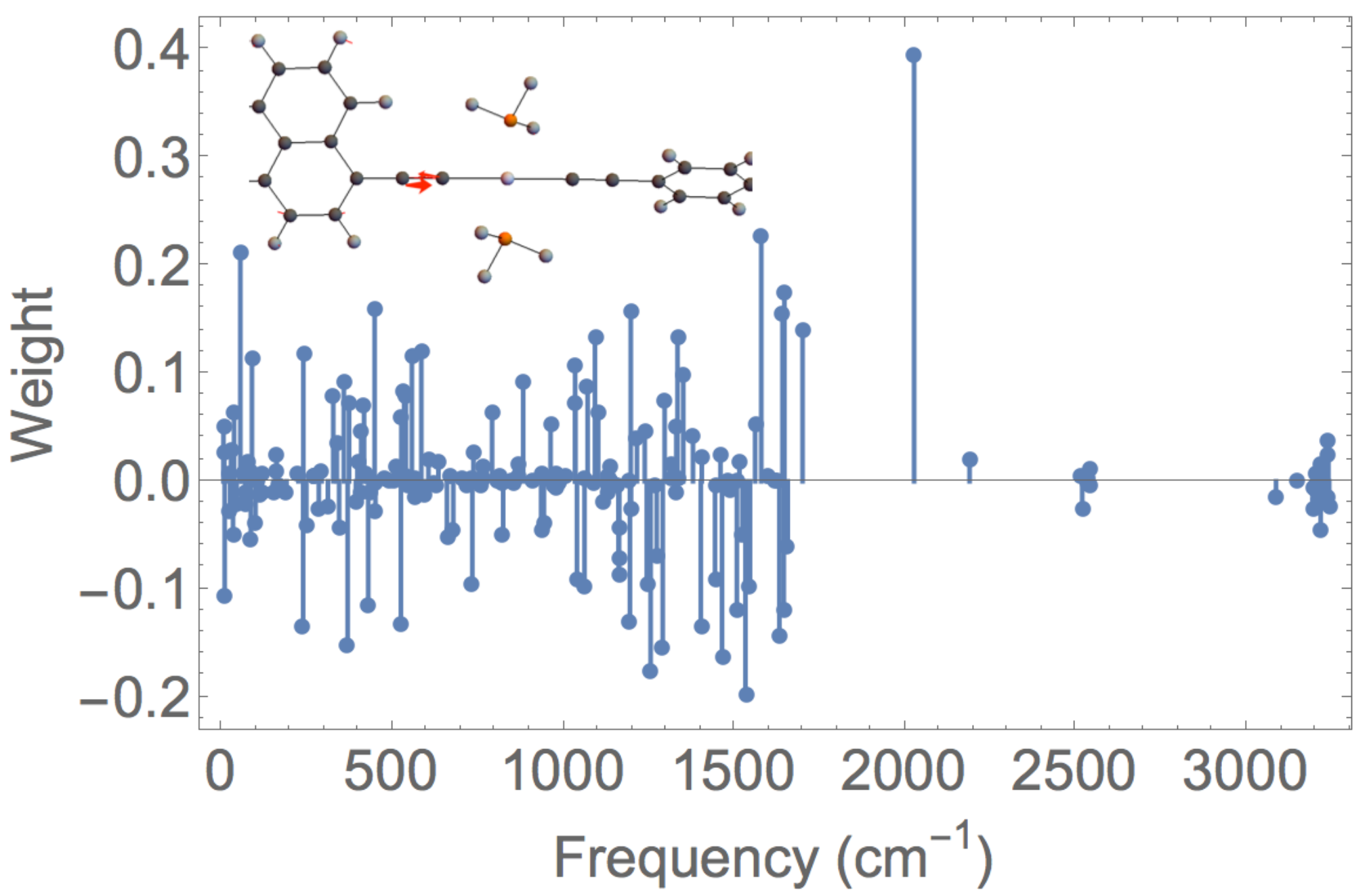}}
\subfigure[]{\includegraphics[width=0.45\columnwidth]{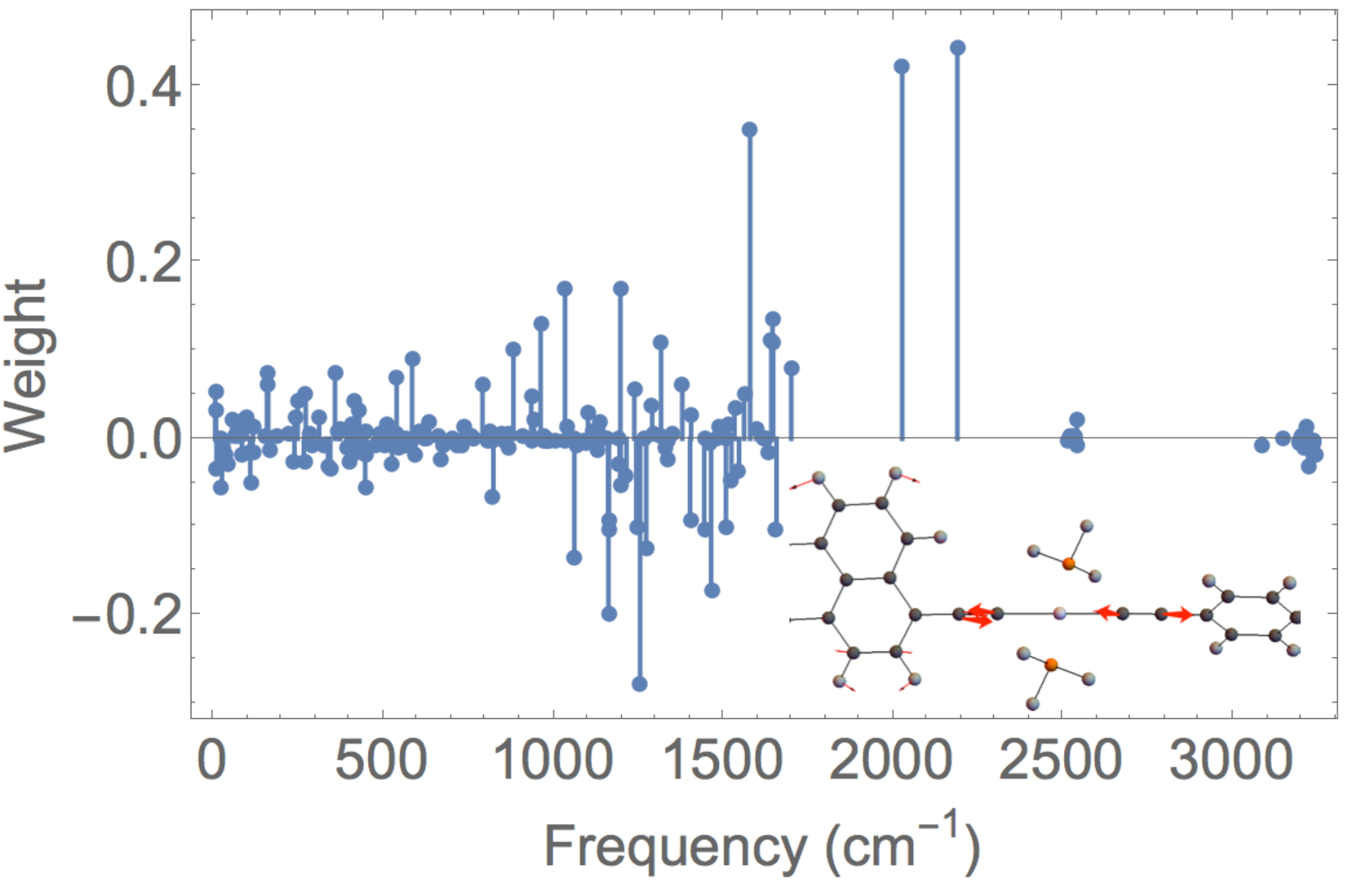}}\\
\subfigure[]{\includegraphics[width=0.45\columnwidth]{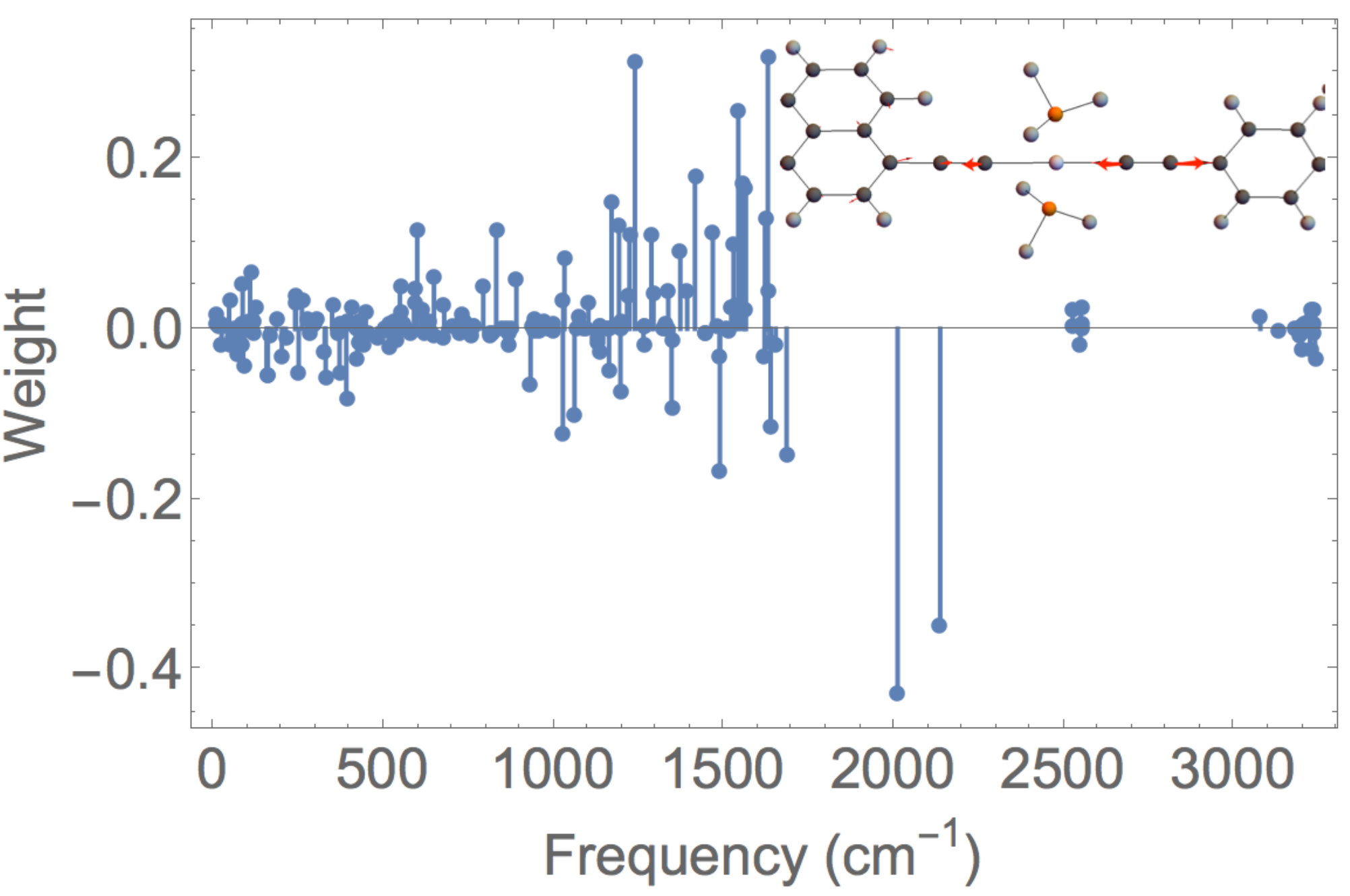}}
\subfigure[]{\includegraphics[width=0.45\columnwidth]{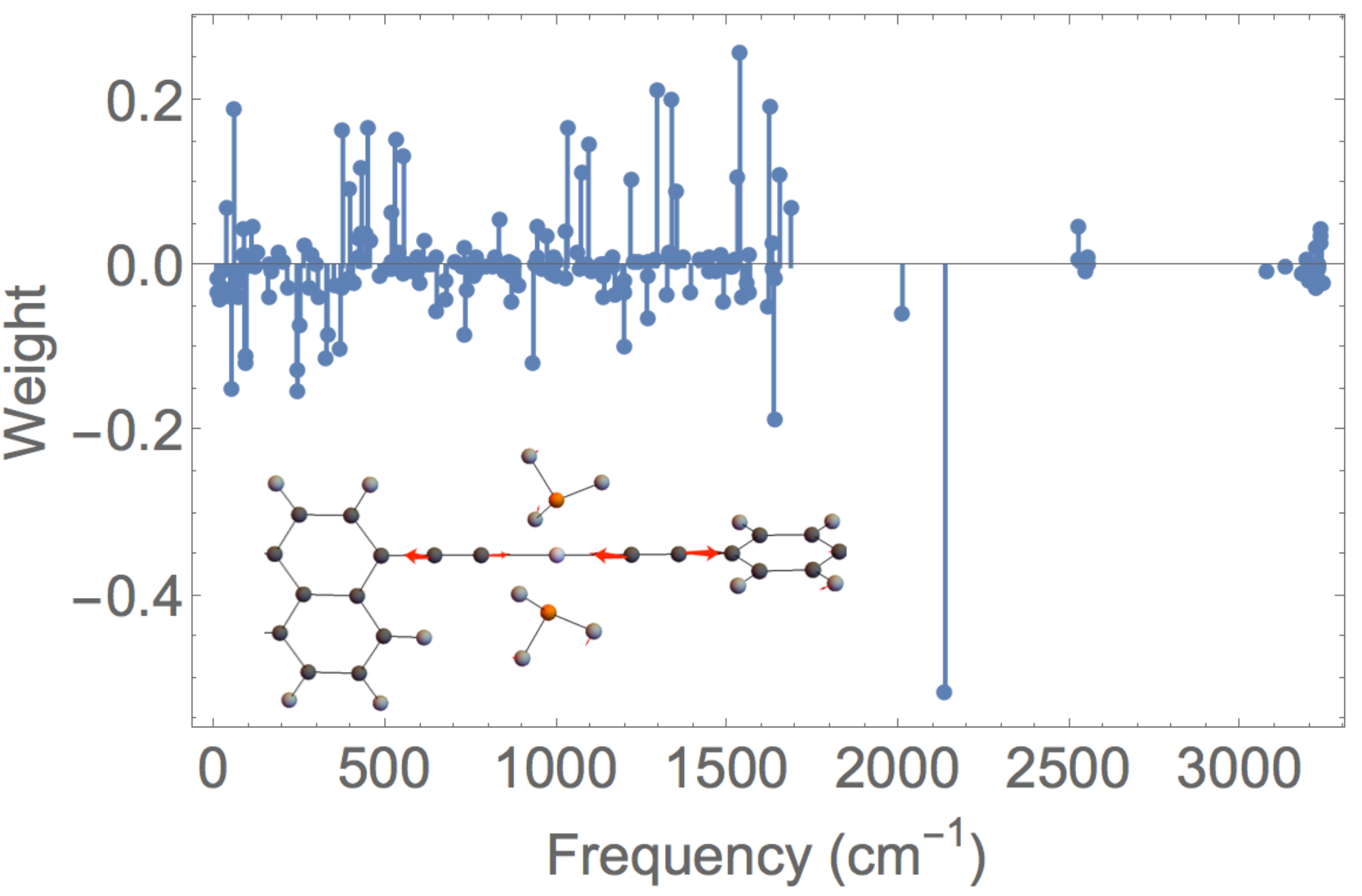}}
\caption{Component projection of the primary mode onto the normal modes for the following transitions:
 (a) CSS $\rightarrow$ $^3$NAP, (b) CT $\rightarrow$ $^3$NAP calculated at $^3$NAP geometry. (c)CT $\rightarrow$ $^3$NAP, and (d) CT $\rightarrow$ CSS calculated at CT geometry.
Embedded molecule shows the atomic displacement vectors of primary mode.
}\label{projT1}
\end{figure*}

Fig. \ref{projT1}(a-d) shows the projection of
the primary mode identified for each transition
onto the normal vibrational modes of the originating state,
{\em i.~e.}, the primary modes calculated at CT geometry
are projected onto the normal modes of the CT state,
 and those at the $^3$NAP geometry
 are projected onto the normal modes of the $^3$NAP state.
In all four cases, the primary mode is dominated by
symmetric and anti-symmetric
contributions from the  C$\equiv$C displacements.
 While both transitions involve acetylene  bond-stretching motions,
the CT$\to$CSS transition involves only the {\em symmetric} combination,
whereas the CT$\to ^3$NAP involves both the {\em symmetric}  and {\em anti-asymmetric} combination.
It is tempting to conclude from this that that the secondary IR push used in the experiments preferentially
excites the anti-symmetric mode and thus selectively enhances the CT$\to ^3$NAP transition.  In fact,
the computed IR oscillator strength of the anti-symmetric mode is an order of magnitude
greater than the symmetric mode.   Similarly, from experiment,
the anti-symmetric normal mode extinction coefficient is 3 times larger than that for the symmetric normal mode.
However,  the time-scale for the IR excitation is sufficient long enough that {\em both} symmetric and
anti-symmetric CC modes are expected to be equally populated by the IR push pulse.

In the CT$\to ^3$NAP transition, both types of acetylene stretching
motions (symmetric and anti-symmetric) contribute more or less
equally to the electronic coupling while in the CT$\to$CSS  transition,
 {\em only} the symmetric acetylene motion carries the majority of the coupling.
This mechanism can be rationalized by the way the vibrational populations
enter into our expression for electron/nuclear coupling correlation function in Eq.~\ref{corrf}.
In principle, the expression was derived assuming a thermal population of the
vibrational modes.  However, if we assume that the role of the IR pulse is to
excite the $C\equiv C$ stretching modes by one vibrational quantum,  then
the value of $\bar n_{i}$  appearing in Eq. ~\ref{corrf}
for those modes should be increased to $\bar n_{i} + 1$.
Consequently, driving these modes with the IR pulse {\em increases} the total
electronic coupling, consistent with
 the experimental observation that IR excitation following formation of the CT states
accelerates the CT$\to ^3$NAP transition relative to the CT$\to$CSS transition.

%Figure 7

\section{Discussion}

We here a review of our work in developing new tools for analysing electronic transitions in 
complex molecular systems.  Central  to our work is the notion that one can systematically 
identify a subset of vibrational modes that capture the majority of the electronic coupling to the 
nuclear motions.  These primary modes capture the short-time dynamics with sufficient accuracy
for computing the salient correlation and response functions necessary for evaluating the
golden-rule rates for state-to-state transitions.   While not a central theme to this review, our 
time-convolutionless master equation method can be used for computing multi-state 
transitions and in cases where the state to state rates are time-dependent.  \cite{tamura2008phonon,pereverzev2009energy}  

We believe that the
 key to understanding and ultimately controlling electron transfer pathways in a complex molecular species
 is through vibronic coupling.   The approach we have delineated in this article offers a systematic
 way to deduce a subset of nuclear motions that are most responsible for driving electronic transitions.
 When paired with the TCLME approach for computing the state-to-state transitions, we can obtain
 rate constants that are in quantitative agreement with experimental rates and probe deeper into the
 dynamics to understand which specific types of nuclear motions are involved in a given transition.
The algorithm illustrated here in the example of photo-induced charge-transfer should be of considerable utility for
understanding of a multitude of light-induced reactions where several electronic states are involved in ultrafast transformations.

\section*{Acknowledgements}
XY thanks Tian Lu for  help with Multiwfn. The work at the University of Houston was funded in part by the
National Science Foundation (CHE-1362006, MRI-1531814)
and the Robert A. Welch Foundation (E-1337).
We thank the Weinstein group at the University of Sheffield for sharing their experimental results  and
many detailed conversations regarding the PTZ-Pt-NAP triad.

%merlin.mbs apsrev4-1.bst 2010-07-25 4.21a (PWD, AO, DPC) hacked
%Control: key (0)
%Control: author (8) initials jnrlst
%Control: editor formatted (1) identically to author
%Control: production of article title (-1) disabled
%Control: page (0) single
%Control: year (1) truncated
%Control: production of eprint (0) enabled
%

%
% ****** End of file apssamp.tex ******

\end{document}